\documentclass[authoryear,preprint,review,12pt,numbers,sort&compress]{elsarticle}
\usepackage{setspace}
\RequirePackage[OT1]{fontenc}
\RequirePackage{amsthm,amsmath}
\RequirePackage[numbers]{natbib}
\usepackage{lineno}
\usepackage{color}
\usepackage{amssymb}
\usepackage{amsmath,algorithm,algorithmic}
\usepackage{amsthm,amsmath,natbib}
\usepackage{epsfig}
\usepackage{graphicx}
\usepackage{rotating}
\usepackage{verbatim}
\usepackage[margin=1in]{geometry}
\newtheorem{thm}{Theorem}

\newtheorem{lem}{Lemma}
\theoremstyle{definition}
\newtheorem{rmk}{Remark}

\journal{SPL}
\begin{document}
\begin{frontmatter}

\title{Phylogenetic Ornstein-Uhlenbeck regression curves}

\author[rvt]{Dwueng-Chwuan Jhwueng}
\address[rvt]{Department of Statistics, Feng-Chia University,
Taichung 40724, Taiwan }

\author[f1,f2]{Vasileios Maroulas\corref{cor1}}
\ead{maroulas@math.utk.edu}
\cortext[cor1]{corresponding author}
\address[f1]{Department of Mathematics, 
The University of Tennessee,
 Knoxville, TN  37966, USA}
 \address[f2]{Department of Mathematical Sciences, 
The University of Bath,
Bath BA27AY, UK}
\begin{abstract}
Regression curves for studying trait relationships are developed herein. The adaptive evolution model is considered an Ornstein-Uhlenbeck system whose parameters are estimated by a novel engagement of generalized least-squares and optimization. Our algorithm is implemented to ecological data.
\end{abstract}

\begin{keyword}
Ornstein-Uhlenbeck processes \sep Generalized linear models: evolutionary regression analysis \sep phylogenetic comparative methods

\MSC[2010] 60H30 \sep 62J12; 62P10

\end{keyword}

\end{frontmatter}


\section{Introduction}\label{intro}
Phylogenetic comparative methods are statistical methods
for analyzing data of groups of related species, called
comparative data in the ecology and evolution literature \citep{beau2012,Buki,Fel85,HansenMartins96}. 

Since the species are related by shared evolutionary history, it
may not be reasonable to view such data as independent,
identically distributed realizations of the same stochastic
process. Instead, information about the shared evolutionary
history, explained by the phylogeny of the species, is often
incorporated into the analysis. A partial list of such
studies is in \citet{Fel04, Har91} and references therein. 

While the phylogenetic tree describes the evolutionary relationship, the trait evolution of $n$ species is considered as the $n$-tuple of random variables which evolves as a continuous Markovian
process whose statistical dependency is described by the phylogenetic tree. For a species, let $y_t$ denote its trait value at
time $t$. A trait value could be the body or brain mass. In this paper, similarly to \cite{Hansenetal08}, we consider
that the response trait evolves toward an optimum, $\theta$. More
precisely, the trait, $y_t$, is a solution of a pertinent
Ornstein-Uhlenbeck (OU) stochastic differential equation (SDE) \citep{Oksendal00} given below,
\vspace{-10pt}
\begin{equation} \label{OUevo1}
dy_t=-\alpha_1(y_t-\theta_t)dt +\sigma_y dW_t^y,\vspace{-10pt}
\end{equation}
where $\alpha_1$ measures the rate of adaptation toward an optimum $\theta$, $W^y_t$ is a white noise with mean zero and appropriate covariance, and $\sigma_y$ is the standard deviation of the random change in the evolutionary process. Considering the OU dynamics of evolution expressed in (\ref{OUevo1}), we note that the deterministic part is responsible for a linearly increasing pull of the trait toward the primary optimum and the stochastic part expresses an indirect change.
Furthermore, we consider that the optimum $\theta_t$ is a solution of an OU stochastic process as well,
\vspace{-10pt}
\begin{equation} \vspace{-10pt}
d\theta_t =-\alpha_2 \theta_t dt +\sigma_\theta dW_t^\theta,  \label{sde optimum}\vspace{-10pt}
\end{equation}
where $\alpha_2$ measures the forces for pulling the optimum process $\theta_t$
back to its own optimum which is assumed to be zero. The noise $W_t^\theta$ is correlated with the noise $W_t^y$ with correlation $\rho$. Because of the presence of the two OU processes which drive the evolution, we call it an OUOU model. \citet{Buki} studied a model where the optimum $\theta$ remained constant and \citet{Hansenetal08} extended the results by studying an OUBM model where the trait evolved according to an OU process and the optimum was a Brownian motion (BM). Considering the optimum as a BM lacks the ability to describe its own central tendency. The novel OUOU model is developed precisely to remedy this matter. 

Based on eqs. \eqref{OUevo1} and \eqref{sde optimum} and taking into account that the optimum $\theta_t$ is linearly dependent on the predictor $x_t$  \citep{Hansenetal08}, we establish an evolutionary regression
curve between the predictor and the response $y_t$ (see Theorem \ref{ThmOUevoreg}). The key step for finding the regression curve is Lemma \ref{reg trait on optimal} which
demonstrates the relationship between the trait values, $y_t$, and
the optimal, $\theta_t$. To estimate the regression parameters, generalized least squares estimates (GLS) \citep{NormanSmith98} are employed. Towards this end, we develop an algorithm that identifies the maximum likelihood estimators using the Powell's method \citep{Pressetal07} which relies on derivative-free univariate line optimization techniques, making the computations feasible. To the best of our knowledge this blending of GLS and Powell's optimization technique  and the corresponding computational scheme are novel in the framework of phylogenetic statistics.

Our methodology is in turn applied to an ecological data set
of the evolution of the woodcreepers (see Figure 2). Woodcreepers use their tail for supporting their body. Adopting the OUOU model, we examine how the tip width (the width of the rachis at the
base of the medial rectrix) would adapt to the base width (the width of
the rachis at the tip of the rectrix). The algorithmic implementation agrees with the overall pattern based on the OUBM model of \citet{Hansenetal08}, however, the OUOU model suggests a better fit for the woodcreepers data set. 

Our paper is organized in the following way. Section \ref{methods} describes the establishment
of the OUOU regression curves. Section \ref{algorithm} provides an
algorithmic implementation for estimating the various parameters
of the regression curves of Section \ref{methods} and Section \ref{data analysis} employs the algorithmic implementation of the OUOU model and applies it in an ecological data set of woodcreepers.
Finally, Section \ref{conclusion} summarizes our findings and suggests future research directions on the topic.


\section{Methodology} \label{methods}
This section displays the theoretical foundation which sets the foundation for estimating the parameters of the regression curve. 

Let the response trait, $y_t$, be a solution of the SDE given in (\ref{OUevo1}). We further assume that the optimum is linearly changing according to the predictor $x$, i.e. $\theta_t = b_0+b_1 x_t$ \citep{Hansenetal08}. Given this linear dependency between $\theta_t$ and $x_t$, the predictor variable $x_t$ is also a solution of an appropriate OU SDE (see Appendix), and $\sigma_\theta = b_1 \sigma_x$, where $\sigma_x$ is the diffusion for the predictor $x$. 
Without loss of generality we assume that the rates of adaptation in (\ref{OUevo1}) and (\ref{sde optimum}), respectively, are equal, i.e. $\alpha_1=\alpha_2=\alpha$. This assumption is not essential, however, it decreases the implementation cost of the algorithm presented in Section \ref{algorithm} since one less parameter is estimated. Below, it is the key lemma where the main theorem of this paper, Theorem \ref{ThmOUevoreg},  is based.

\begin{lem} \label{reg trait on optimal}
Let consider that the trait, $y_t$, and its optimum, $\theta_t$, evolve via the OU dynamics described in eqs. (\ref{OUevo1}) and (\ref{sde optimum}), with initial conditions $y_0$ and $\theta_0$ (trait values at the root), respectively, and equal adaptation rates ($\alpha_1=\alpha_2=\alpha$). The regression of the trait on the
optimum is given by 
\vspace{-10pt}
\begin{equation}
{E}[y_t|\theta_t] = \hat{\beta}_0(t) +
\hat{\beta}_1(t) \theta_t, \label{fullreg}
\end{equation}
\vspace{-15pt} where \vspace{-5pt}
\begin{align}
\hat{\beta}_1(t) &= \frac{(1+2\rho)(1-\exp(-2\alpha t)) - 2\alpha t \exp(-2\alpha t)}{2(1-\exp(-2 \alpha t))}, \label{b1} \\
\hat{\beta}_0(t) &= \alpha\theta_0 t \exp(-\alpha t) + y_0
\exp(-\alpha t) - \hat{\beta}_1(t)\exp(-\alpha t) \theta_0. \label{b0} 
\end{align}
\end{lem}
\begin{proof}
Given the pair of dynamics of the trait and its corresponding optimum, $Z_t=(y_t, \theta_t)^T$, where $T$ denotes transpose, we can summarize the eqs. (\ref{OUevo1}) and (\ref{sde optimum}) in a multivariate OU process \begin{equation} \label{sOU} dZ_t= A Z_t dt + D dW_t, \end{equation}
where 
$A= \begin{pmatrix}
-\alpha & \alpha  \\
0 & -\alpha \\
\end{pmatrix}$, $D$ is diffusion matrix satisfying $DD^T= \begin{pmatrix}
\sigma_y^2& \rho \sigma_y \sigma_\theta  \\
\rho \sigma_y \sigma_\theta & \sigma_\theta^2 \\
\end{pmatrix}$ and $W_t$ is a bivariate Wiener process. 
Employing arguments of the multivariate OU processes, see eg. \citet[p. 109]{Gardiner04}, the solution of eq. \eqref{sOU} is given by the following equation, 
\vspace{-10pt}
\begin{equation} \label{ssOU}
Z_t=Z_0 \exp(-At) + \int_0^t \exp(-A(t-s)) D dW_s.\vspace{-10pt}
\end{equation} 
The solution of the OU system based on the expression in eq. \eqref{ssOU} yields the mean, the variance of the trait and its optimum, respectively, and their covariance. The reader may refer to \citet[Section 4.4.4.]{Gardiner04} for a detailed treatment on the subject.
\vspace{-10pt}
\begin{align} 
\mathbb{E}[\theta_t]&=\theta_0 \exp(-\alpha t)  \label{theta1} \\
\mbox{Var}[\theta_t] &=\sigma_\theta^2\frac{1-\exp(-2\alpha t)}{2\alpha}  \label{vt1}\\
 \mathbb{E}[y_t] &=\alpha\theta_0 t \exp(-\alpha t) + y_0 \exp(-\alpha t) \label{y1} \\
 \mbox{Cov}[y_t,\theta_t] &= \sigma_\theta^2\{\frac{ (1+2\rho)(1-\exp(-2\alpha t))}{4\alpha} - \frac{t}{2}\exp(-2\alpha t)\} \label{covyt}\\
 \mbox{Var}[y_t] &= (\frac{\sigma_y^2}{2\alpha}+\frac{(1+2\rho)\sigma_\theta^2}{4\alpha})(1-\exp(-2\alpha t))-\sigma_\theta^2\frac{(1+2\rho+\alpha t)t}{2}\exp(-2\alpha t) \label{vy1}
\end{align}
Considering the estimates of the involved parameter in eqs. \eqref{theta1}-\eqref{covyt} and based on standard regression arguments, the result follows. 
\end{proof}
The central theorem of this paper, presented below, establishes the evolutionary regression curve when the trait and the predictor evolve according to OU dynamics.
\begin{thm}(Evolutionary Regression Curve) \label{ThmOUevoreg}
Let consider that the trait value $y_t$ and its optimum are solutions of the OU SDEs (\ref{OUevo1}) and (\ref{sde optimum}), respectively. Furthermore, the optimum process $\theta_t$ is related to the predictor $x_t$ via a linear regression of the form, $\theta_t=b_0+b_1x_t$. Then the evolutionary regression curve of the trait $y_t$ on $x_t$ is given by,
\vspace{-10pt}
\begin{equation}
\mathbb{E}(y_t|x_t) = \hat{\beta_0}(t) + b_0 \hat{\beta_1}(t) + b_1 \hat{\beta_1}(t)x_t,
\label{pi}\vspace{-10pt}
\end{equation}
\noindent  where the parameters, $\hat{\beta_0}(t)$ and $\hat{\beta_1}(t)$, are defined in Lemma \ref{reg trait on optimal}.
\end{thm}
\begin{proof}
By Lemma \ref{reg trait on optimal} the regression line is $\mathbb{E}[y_t|\theta_t] = \hat{\beta}_0(t) +
\hat{\beta}_1(t) \theta_t$. Since the optimum $\theta_t$ depends linearly on 
$x_t$, i.e. $b_0+b_1x_t$, the result yields. 
\end{proof}
Next, let consider a species whose pair of the ancestral trait and its corresponding optimum is $(y_a,\theta_a)$ at time $t_a$. The time $t_a$ can be regarded as the branch length (evolutionary time) from the root of the phylogeny to the most recent common ancestor of the  species $i$ and $j$. Also, let assume that the speciation occurs at time $t_a$, and it takes $t_i, \; t_j$ time  for species $i, \; j,$ respectively to evolve until the current time. It is considered herein that $t_i=t_j=\frac{t_{ij}}{2}$, where $t_{ij}=t_i+t_j$ (see Figure 1). 

In what follows, the time index $t$ is the evolutionary time $t_{ij}$ for the phylogenetic data analysis. Furthermore, for the sake of presentation, we suppress the time index unless otherwise.
\begin{figure} \label{fig1}
\vspace{-10pt}
\begin{center}
\includegraphics[width=4cm,height=4cm,angle=-90]{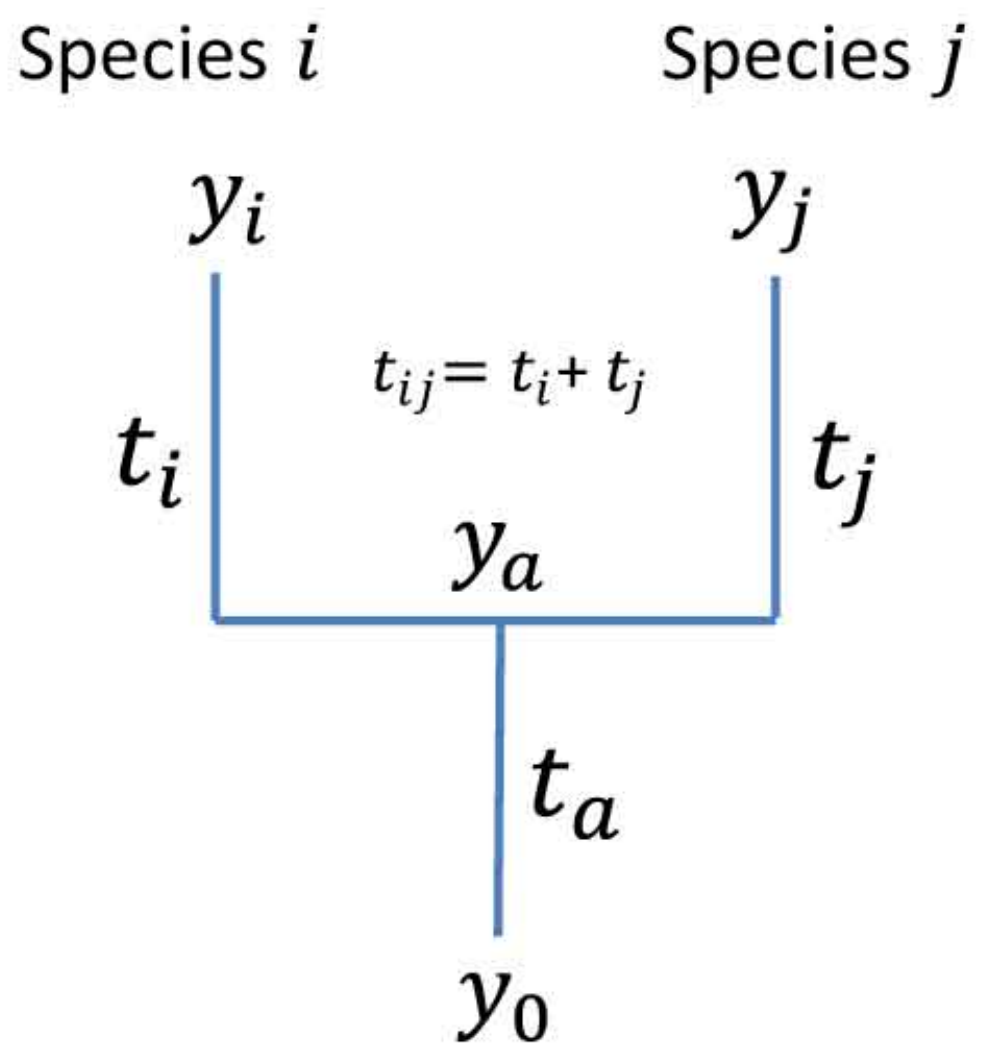}
\end{center}
\vspace{-15pt}
\caption{Evolutionary scheme for two species $i$ and $j$ with traits $y_i$ and $y_j$, respectively.}
\end{figure}
Based on eq. \eqref{y1}, the conditional expectation of the trait $y_i$ given its ancestral value $y_a$ at time $t=t_a$, $\mathbb{E}[y_i|y_a]$, equals to
\vspace{-10pt}
\begin{equation} \label{cea}
\mathbb{E}[y_i|y_a] =\alpha \theta_a \frac{t_{ij}}{2} \exp(-\alpha t_{ij}/2) +y_a\exp(-\alpha
t_{ij}/2).\vspace{-10pt}
\end{equation} 
According to eq. \eqref{cea} and the result of \citet{HansenMartins96} that the covariance between traits is given by $ Cov[y_i,y_j] = Cov[[\mathbb{E}[y_i|y_a],\mathbb{E}[y_j|y_a]],$ one can deduce that the covariance between two species $y_i, y_j$ with a common ancestor $y_a$ equals to 
\begin{align}  
Cov[y_i,y_j] =Var [\frac{\alpha t_{ij}}{2}\exp(-\alpha t_{ij}/2) \theta_a
+ \exp(-\alpha t_{ij}/2)y_a] \notag \label{cov of traits}\\ = \frac{\alpha^2 t_{ij}^2}{4} \exp(-\alpha t_{ij}) Var[\theta_a] + \exp(-\alpha t_{ij}) Var[y_a]  +\alpha t_{ij}\exp(-\alpha t_{ij})Cov[y_a,\theta_a],
\end{align}
where  $Var[\theta_a]$, $Cov[y_a,\theta_a]$ and $Var[y_a]$ are determined by eqs. (\ref{vt1}), (\ref{covyt}) and (\ref{vy1}), respectively, by replacing $t$ with $t_a$. Our aim is to estimate the regression parameters $\hat{\beta}_1, \hat{\beta}_0$ and $b_0,b_1$ since they are involved in the regression of Theorem \ref{ThmOUevoreg}. Generalized least squares methods are used and thus we need to compute the variance and the covariance of the
residuals. Let denote $r_i = y_i - \mathbb{E}[y_i| \theta_i]$ to be the
$i$th residual associated with the $i$th prediction from the
regression curve of Lemma \ref{reg trait on
optimal}. The covariance between a pair of residuals, $Cov(r_i,r_j)$, equals to
\begin{align} 
v_{ij} = Cov[r_i,r_j]&=Cov[y_i-\mathbb{E}[y_i|\theta_i],y_j-\mathbb{E}[y_j|\theta_j]] \label{rescov} \\
&=Cov[y_i,y_j]-Cov[y_i,\mathbb{E}[y_j|\theta_j]] -Cov[y_j,\mathbb{E}[y_i|\theta_i]]
+Cov[\mathbb{E}[y_i|\theta_i],\mathbb{E}[y_j|\theta_j]],\notag 
\end{align} 
where the covariance $Cov[y_i,y_j]$ between two traits is given in eq. $(\ref{cov of traits})$. Based on  eq. \eqref{fullreg}, the rest of covariances in eq. (\ref{rescov}) are equal to, $Cov[y_i,\mathbb{E}[y_j|\theta_j]]= \hat{\beta_1}(t_j)Cov[y_a,\theta_a]$ and 
$Cov[\mathbb{E}[y_i|\theta_i],\mathbb{E}[y_j|\theta_j]] =
\hat{\beta}_1(t_i)\hat{\beta}_1(t_j)Var[\theta_a]$,
where $\hat{\beta}_1(t_i)$ and $\hat{\beta}_1(t_j)$  are the regression slopes of eq. (\ref{b1}) by replacing the time $t$ with $t_i$, and $t_j$, respectively. 
%
%
 %
\section{The algorithm} \label{algorithm}
The construction of the algorithm while adopting the framework of Section \ref{methods} is presented in this section. Our algorithm with input comparative data and a phylogenetic tree establishes the regression curve of Theorem \ref{ThmOUevoreg} by identifying the maximum likelihood estimators (MLEs) for 
the adaptation rate $\hat{\alpha}$, the variances $\hat{\sigma}^2_y$ and the regressors' vector $\hat{b}$ via a novel engagement of GLS and Powell's method \citep{Pressetal07}. 

Let $\underline{x}=(x_1,x_2,\cdots,x_n)^T$ be the observed value of the predictor variable for the $n$ species. 
Consider the model $y= Xb + \epsilon$, where the response variable $y \in \mathbb{R}^{n \times 1}$ represents the mean trait value of species, $X \in \mathbb{R}^{n \times 2}$ is the design matrix such that $X=(\textbf{1}, \underline{x}),$ and $\textbf{1} \in \mathbb{R}^{n\times 1}$ is the column vector with one in its entries and $\underline{x}$ represents the predictor variable after a pertinent phylogenetic correction, and $b=(b_0,b_1)^T$ the regressors' vector. Furthermore, the error structure $\epsilon$ follows a normal distribution, $\epsilon \sim N(0,V),$ where each entry $v_{ij}$ in $V$ equals to $Cov[r_i,r_j]$ in eq. (\ref{rescov}). 
We observe that the covariance matrix $V$ depends on the parameter $b_1$ and further the square of the diffusion coefficient of the optimum $\sigma_\theta^2=b_1^2\sigma_x^2$. Therefore we do not have a closed form for the estimates. Consequently we need to estimate first the mean and variance-covariance of the predictor $x$ and then we can use the regular GLS estimates given by 
\begin{equation}\label{GLS}
\hat{b}=(X^TV^{-1}X)^{-1}X^TV^{-1}y
\end{equation}
in order to approximate $b$.
Since the trait of each species is a continuous Markov process (an OU process in our case), the covariance between the trait for a pair of species can be derived by incorporating the time and the nature of the Markov process. Relying on an OU process, the covariance between two predictors $x_i,x_j$ enjoys the closed form 
\begin{equation} \label{eq:cov_x}
Cov[x_i,x_j]=\sigma_x^2A_\alpha[i,j]=\sigma_x^2\left[\exp(-\alpha t_{ij})\frac{1-\exp(-2\alpha t_a)}{2\alpha} \right].
\end{equation}
The derivation of the covariance is similar to \cite{Buki} and \cite{Jhwueng13} and thus it is delegated to the Appendix. Note that the matrix $A_\alpha$ can be evaluated directly when the phylogenetic tree with known branch lengths is given \citep{Buki, beau2012}. In addition, according to  \citet{Jhwueng13}, the estimated mean, $\hat{\mu}_x,$ and the corresponding estimated variance, $\hat{\sigma}^2_x,$ of the predictor $x$ depend on $\alpha$ and are as follows
\begin{equation}
\hat{\mu}_x = (\textbf{1}^T A_{\alpha} \textbf{1})^{-1} \textbf{1}^T A_{\alpha}^{-1} \underline{x} \text{ and } 
\hat{\sigma}_x^2 = \frac{(\underline{x}-\hat{\mu}_x\textbf{1})^T A_{\alpha}^{-1}(\underline{x}- \hat{\mu}_x\textbf{1})}{n}, 
\label{mles}
\end{equation}
Since the optimum is linearly dependent with the observed predictor $x_t$, i.e. $\theta_t=b_0+b_1x_t$, it follows immediately that  $\sigma^2_\theta=b_1^2 \sigma_x^2$. Therefore the remaining parameters to be estimated in the covariance matrix $V$ are the regressors' vector $b$, the adaptation rate $\alpha$, the correlation $\rho$, and the variance of the traits $\sigma_y^2$. However the GLS of eq. \eqref{GLS} is not feasible to work since it depends on $V$. Thus, we initialize our algorithm incorporating an ordinary least-squares (OLS) estimate instead, $\hat{b}^0=(X^TX)^{-1}{X}^Ty$ which is independent of the covariance matrix $V$. Subsequently, plugging $\hat{b}^0$ into the log-likelihood function, $\ell$, defined below in eq. \eqref{lhood}, 
 \begin{equation} \label{lhood}
\ell(b, \alpha, \sigma^2_y,\rho)=-\frac{n}{2}\log (2\pi) -\frac{1}{2}\log(\det(V)) -\frac{1}{2}(y-Xb)V^{-1}(y-Xb),
\end{equation}
the MLE triplet $(\hat{\alpha},\hat{\sigma}^2_y,\hat{\rho})$ is estimated by optimizing the log-likelihood $\ell$ on an appropriate domain for $(\alpha, \sigma_y^2, \rho)$ through Powell's method. The constraints of the optimization are weak, i.e. the upper bound for $\alpha > 0$ is arbitrary, and the variation of data should not exceed a range as much as the difference between the maximum and the minimum observation, in other words we set the domain of $\sigma_y^2$ to  be $[0,y_{(n)}-y_{(1)}]$ where $y_{(n)}$ and $y_{(1)}$ are the largest and smallest value of observation, respectively, and  the domain for correlation $\rho$ is by definition $[-1,1]$. After establishing the MLEs ($\hat{\alpha},\hat{\sigma}_y,\hat{\rho})$, we estimate the regressor $\hat{b}$ by eq. \eqref{GLS}. The process is iterative until the discrepancy, $|| \hat{b} - \hat{b^0}||$, is less than some threshold $\delta$. Once the estimation converges, the evolutionary regression curve is propagated according to eq. (\ref{pi}). Our computational implementation is summarized in Algorithm 1.
\begin{rmk}
Note that given the observed response traits $y=(y_1,y_2,\cdots,y_n)^T$, the basal ancestral species is the fixed starting point of the phylogenetic tree. Thus we can estimate $y_0$ and $\theta_0$ by
$y_0=\theta_0=(\textbf{1}^T A_{\alpha} \textbf{1})^{-1} \textbf{1}^T A_{\alpha}^{-1} y$. Therefore, the estimates of $\hat{\beta}_1, \hat{\beta}_0$ in eqs. \eqref{b1} and \eqref{b0}, respectively, are searched after the MLE of $\alpha$ is found. 
\end{rmk}

\begin{algorithm}\label{alg}
\caption{Estimation of the phylogenetic evolutionary regression curve} \small
\begin{algorithmic}[1]
\REQUIRE
Phylogenetic tree with known branch length and comparative data.
\ENSURE Phylogenetic evolutionary regression curve under OUOU model.
\itemsep 0.05cm
    \STATE Given data $y$, set $\hat{b}^0=(\hat{b}_0^0,\hat{b}_1^0)^T=(X^TX)^{-1}X^Ty$;
     \STATE 
     Given data $\underline{x}$, compute the corresponding MLEs of predictor as functions of $\alpha$ relying on eq. (\ref{mles}). 
     \STATE Use the estimates $\hat{\mu}_x,\hat{\sigma}_x^2, \hat{b}^0$ 
     as an input for the covariance matrix $V$ in eq. (\ref{rescov}). The log-likelihood function $\ell$ of eq. \eqref{lhood} depends only on $\alpha,\sigma^2_y$, and $\rho$; 
            \STATE Search the MLEs $(\hat{\alpha}, \hat{\sigma}_y^2, \hat{\rho})$ by optimizing $\ell(\alpha,\sigma_y^2,\rho)$ through Powell's method;
            \STATE Incorporate the MLEs into the design matrix and covariance matrix to obtain $\hat{X}$ and $\hat{V}$ and approximate the regression estimate by $\hat{b}=(\hat{X}^T\hat{V}^{-1}\hat{X})^{-1}\hat{X}^T\hat{V}^{-1}y$;
          \STATE Set  $\delta=||\hat{b}-\hat{b}^0||$;          
\IF {$\delta < err$}
\RETURN $\hat{b}$ and $(\hat{\alpha}, \hat{\sigma}_y^2, \hat{\rho})$;
\ELSE
\STATE Set  $\hat{b}^0=\hat{b}$ and go back to Step 2;
\ENDIF
\STATE Use MLEs ($\hat{\alpha},\hat{\rho}$) and optimal regression estimates $\hat{b}$ of Step 7 to obtain the phylogenetic evolutionary regression curve $\mathbb{E}[y_t|x_t]= \hat{\beta}_0 +\hat{b}_0\hat{\beta}_1+ \hat{b}_1\hat{\beta}_1 x_t$ given in Theorem \ref{ThmOUevoreg}. 
    \normalsize
\end{algorithmic}
\end{algorithm}
\begin{figure} \label{fig2}
\vspace{-50pt}
\centering
\includegraphics[scale=0.40,angle=00]{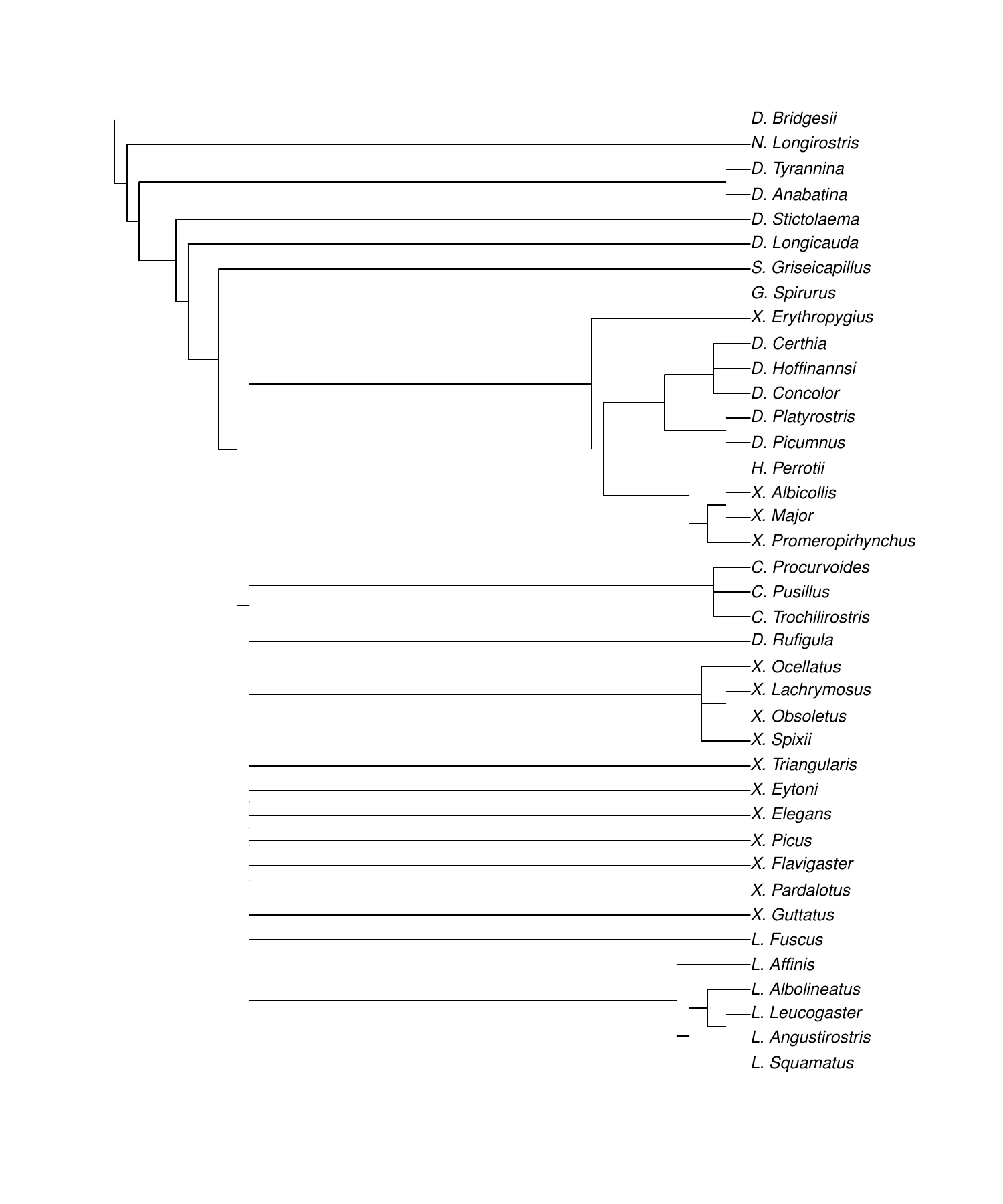}
\vspace{-50pt}
\caption{The phylogenetic tree of the 39 woodcreeper species as in \citet{Tubaro2002} using the algorithm of construction of a phylogenetic tree from \citet{Paradis04}. }
\end{figure}
\begin{rmk}
This algorithm is different from the approach in \citet{Hansenetal08} where the regressors $\hat{b}$ are first estimated though GLS. Then $\hat{b}$ is entered into the covariance matrix $V$. The search of MLE $(\hat{\alpha},\hat{\sigma}^2_y)$ for the log-likelihood function $\ell$ is conducted under the grid base method which is an exhaustive search through a manually specified subset of the hyperparameter space of a learning algorithm. It is very computational expensive for a higher dimensional search \citep{berg2012}. Powell's method in contrast searches MLEs based on the notion of conjugate directions without considering derivatives and thus the function needs not to be differentiable.
\end{rmk}
\subsection{Data Analysis} \label{data analysis}
Using Algorithm 1 of the OUOU model, we establish the evolutionary regression curve for the woodcreepers data set of $n=39$ species \citep{Tubaro2002}.  The explanatory variable $x$ is considered the rachis width at the base and the corresponding response, $y$, the rachis width at the tip.
For this particular problem, the predictor and the response variables do not necessarily evolve in a correlated environment and thus the noise correlation, $\rho$, is set equal to 0.
Furthermore, note that the predictor indirectly influences the trait through its influence on the fitness optimum $\theta$ of the response which evolves under the OU scheme.
We set the threshold for the error, $err$, not to exceed $ \delta = 10^{-5}$ (Step 6 in Algorithm1) for identifying the convergent regressors. The regression output and the results from the regression comparison are given in the Table \ref{tab:RegressionOutput}. Both OUBM of \citep{Hansenetal08} and our OUOU model suggest a similar increasing pattern between the tip and the base width. On the other hand, the values of $r^2$ and AICc of our method suggest a better fit for the data. Figure 3 displays the two evolutionary curves along with the data. The \textbf{R} source code and data set are available in www.tonyjhwueng.info/OUreg. 
\begin{table*}
	\centering
		\begin{tabular}{|c|c|c|c|c|}
			\hline Model & Method & Regression Line & $r^2$ & AICc  \\ \hline
			M1: OUOU & Sections \ref{methods} \& \ref{algorithm} & \textbf{$y=0.08+0.24x$} & 23.68\% & -31.83 \\  \hline
			M2: OUBM & \citet{Hansenetal08} &  \textbf{$y=0.06+0.28x$} & 22.19\% & -32.72 \\ \hline
		\end{tabular}
	\caption{Comparison of regression curves}
	\label{tab:RegressionOutput}
\end{table*}

\begin{figure} \label{fig3}
\centering
\includegraphics[scale=0.30]{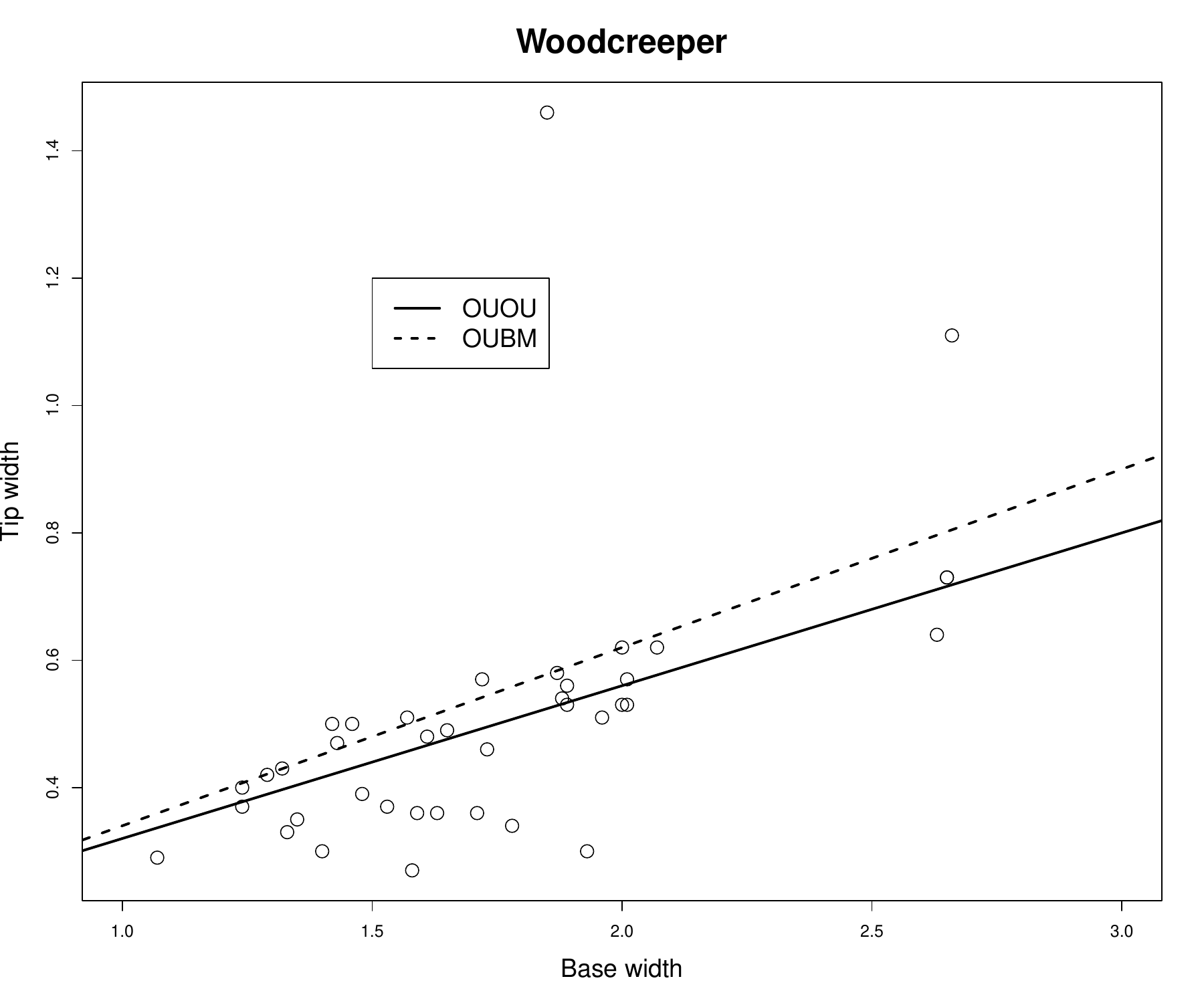}
\vspace{-15pt}
\caption{Evolutionary regression curves for OUBM model (dash line) and OUOU (solid line).}
\end{figure}
%
%
%
\section{Conclusion} \label{conclusion}
A novel approach to the phylogenetic regression analysis was presented by considering that a trait of species evolved according to OU dynamics and the trait's optimum also propagated according to a pertinent OU SDE. The OUOU model generalizes the models studied by \citet{Buki} and \citet{Hansenetal08} and it was implemented algorithmically using a novel blending in phylogenetic statistics of Powell's optimization technique and generalized least squares methods. This algorithm was then applied to a real ecological data set. Our results yielded a better fit than the existing approach. A further research direction is that one could focus on the rate of adaptation and assumes that it also evolves as a continuous Markov process. In this case a highly correlated system of three equations, one for the evolution of the trait, one for its optimum and one for the rate of adaptation should be considered. Furthermore, the OUOU model treated the optimum of the response trait as a linear combination with the predictor. It is interesting to examine whether there exists in some cases a meaningful nonlinear relationship and if it yields accurate results. We will investigate such models in the future, however, it is of paramount importance to maintain the algorithmic complexity of these models when their implementation is considered.
%
%
\section*{Appendix}
The variance-covariance structure for predictor $\underline{x}=(x_1,x_2,\cdots,x_n)^T$ given in eq. \eqref{eq:cov_x} is derived herein. The optimum $\theta_t$ is the solution of the OU-SDE, given in  eq. (\ref{sde optimum}). Since  $\theta_t=b_0+b_1x_t$, it implies that the predictor satisfies an appropriate OU SDE as well. 
The constant term $\frac{b_0}{b_1}$ does not affect the computation of the variance covariance structure of the predictor $x$, and for ease of presentation, one may easily deduce that $x_t$ is the solution of the OU-SDE
$dx_t= -\alpha x_tdt+\sigma_xdW_t^x,$ where $\sigma_x$ is the diffusion parameter that measures the intensity of the random fluctuation in the evolutionary process. Assuming that $x(0)=x_0$ at the initial time $t=0$, the solution is given by $x_t=x_0\exp(-\alpha t) + \sigma_x \exp(-\alpha t)\int_0^t\exp(\alpha s)dW^x_s $ and its corresponding expected value is $E[x_t|x(0)=x_0]=x_0\exp(-\alpha t)$. Let suppress the conditionals for the sake of presentation, i.e. we write $E[x_t]= E[x_t|x(0)=x_0]$, and determine the covariance between a pair of species $x_i$ and $x_j$ of the $i$th and $j$th species, respectively, which are solutions of the OU-SDE. Given the phylogeny as shown in Figure 1, the covariance for $x_i$ and $x_j, \; Cov[x_i(t_i+t_a),x_j(t_j+t_a)]=Cov[x_i(t_i+t_a),x_j(t_j+t_a)|x_i(0)=x_j(0)=x_0],$ is derived \vspace{-10pt}
\begin{align}
&E[x_i(t_i+t_a) x_j(t_j+t_a)] -E[x_i(t_i+t_a)] E[x_j(t_j+t_a)] \notag \\ &=\sigma_x^2\exp(-\alpha(t_{ij}+t_a))E\Bigl[ \bigl(\int_0^{t_i+t_a}e^{\alpha s}dW_s^x \bigr) \bigl(
\int_0^{t_j+t_a}e^{\alpha s}dW_s^x \bigr) \Bigr] \notag \\
&=K\cdot E\left[ \left(\int_0^{t_a}e^{\alpha s}dW_s^x + \int_{t_a}^{t_i+t_a}e^{\alpha s}dW_s^x\right) \cdot 
\left(\int_0^{t_a}e^{\alpha s}dW_s^x + \int_{t_a}^{t_j+t_a}e^{\alpha s}dW_s^x\right) \right], \vspace{-10pt}
\label{Covij}
\end{align}
where $K=\sigma_x^2\exp(-\alpha(t_{ij}+t_a))$ and $t_{ij}=t_i+t_j$. However, the traits evolved independently after time $t_a$ and thus $E\left[\int_{t_a}^{t_a+t_i}e^{\alpha s}dW_s^x \cdot \int_{t_a}^{t_a+t_j}e^{\alpha s}dW_s^x\right]=0$. Moreover, due to independence and properties of It\={o} integrals during successive increments, $E\left[\int_0^{t_a}e^{\alpha s}dW_s^x \cdot \int_{t_a}^{t_a+t_i}e^{\alpha s}dW_s^x\right]=E\left[\int_0^{t_a}e^{\alpha s}dW_s^x \cdot \int_{t_a}^{t_a+t_j}e^{\alpha s}dW_s^x\right]=0$. Therefore, taking into consideration these arguments and It\={o} isometry, the covariance expressed in eq. \eqref{Covij} equals to:
$Cov[x_i,x_j] =  \frac{\sigma^2_x}{2\alpha}\exp(-\alpha(t_i+t_j+2t_a))(\exp(2\alpha t_a)-1)= \frac{\sigma^2_x}{2\alpha}\exp(-\alpha t_{ij})(1-\exp(-2\alpha t_a)) = \sigma_x^2 A_\alpha[i,j] .$
%

\section*{Acknowledgements}
We would like to thank one anonymous Associate Editor and one anonymous reviewer for their comments which allowed us to improve our manuscript substantially. The first author's work was partially supported by the National Science Council Award $\#$NSC-101-2118-M-035-002, Taiwan, ROC. The second author's work was partially supported by the Simons Foundation Award $\#$279870, the Leverhulme Trust Fellowship Award $\#$RE-MA1097 and the University of Tennessee. Both authors' work was also supported by the National Institute for Mathematical and Biological Synthesis (NIMBioS), an Institute sponsored by the National Science Foundation, the U.S. Department of Homeland Security, and the U.S. Department of Agriculture through NSF (National Science Foundation) Award $\#$EF-0832858.


\bibliographystyle{model2-names}
\bibliography{ref}

\end{document}